\documentclass{IEEEtran}
\usepackage{cite}
\usepackage{amsmath,amssymb,amsfonts}
\usepackage{graphicx}
\usepackage{textcomp}
\usepackage{booktabs}
\usepackage{array}
\usepackage{multirow}
\usepackage{subfig}
\usepackage{nicefrac}
\def\BibTeX{{\rm B\kern-.05em{\sc i\kern-.025em b}\kern-.08em
    T\kern-.1667em\lower.7ex\hbox{E}\kern-.125emX}}
\begin{document}
\title{Stability of Charge Collection Efficiency in a Novel
Graphene-Optimized Silicon Carbide Detector Under 160 keV X-Ray Irradiation}
\author{{Yingjie. Huang, Congcong. Wang,  Jingxuan. He, Yi. Zhan, Zhenyu. Jiang, Xiyuan. Zhang and Xin. Shi}
\IEEEmembership{}
\thanks{This work is supported by the National Natural Science Foundation of China (Nos. 12305207, 12375184 and 12405219) and support from CERN DRD3 Collaboration. (Corresponding author: Congcong Wang)}
\thanks{Jingxuan He is with the Institute of High Energy Physics, Chinese Academy of Sciences, Beijing
100049, China, and also with jilin University, Changchun, 130012, China.}
\thanks{Congcong Wang is with
the Institute of High Energy Physics, Chinese Academy of Sciences, Beijing 100049, China, and also with the State Key Laboratory of Particle Detection and Electronics, Beijing 100049, China. (e-mail: wangcc@ihep.ac.cn)}
\thanks{ Jingxuan He and Yi Zhan are with Qufu Normal University, Rizhao, 276827, China.}
\thanks{Zhenyu Jiang is with the Institute of High Energy Physics, Chinese Academy of Sciences, Beijing
100049, China, and also with Liaoning University, Liaoning 110136, China.}
\thanks{Xiyuan Zhang and Xin Shi are with the Institute of High Energy Physics, Chinese Academy of Sciences, Beijing 100049, China.}}
\maketitle
\begin{abstract}
A novel graphene-optimized silicon carbide PIN detector was fabricated. Its electrical properties, charge collection performance and signal rise time were evaluated under non‑irradiated conditions and under X‑ray irradiation with an energy of 160 keV at doses of 0.1 MGy and 1 MGy. The leakage currents of the detectors under non‑irradiated, 0.1 MGy, and 1 MGy irradiation conditions are approximately $1.45 \times 10^{-10}$~A, $1.51 \times 10^{-10}$~A, and $1.57 \times 10^{-10}$~A, respectively. The effective doping concentration of the detector is approximately $8.08 \times 10^{13}$ cm$^{-3}$ before and after irradiation, with no significant change. The rise times of the signals from $\alpha$ particles signal detected by the detector under unirradiated, 0.1 MGy, and 1 MGy X-ray irradiation conditions are 336 ps, 368 ps, and 387 ps, respectively. The rise times of the $\beta$ particles signal detected by the detector under unirradiated, 0.1 MGy, and 1 MGy X-ray irradiation conditions are 342~ps, 375~ps, and 398~ps, respectively. After 0.1 MGy and 1 MGy X‑ray irradiation, the charge collection efficiencies (CCEs) of the detector for $\alpha$ particles are 97.2\% and 90.0\%, respectively; for $\beta$ particles, they are 100.0\% and 97.0\%, respectively. Experiments confirm that 160 keV X-ray irradiation may not cause significant displacement damage in the 4H-SiC, and the minor performance degradation may be attributed to ionization-induced changes in the graphene electrode.
The detector exhibits excellent charge collection performance and fast time response. These results demonstrate stable performance under extreme X-ray exposure, highlighting the detectors potential for radiation-hard applications in high-energy physics, space missions, and nuclear reactor monitoring.

\end{abstract}
\section{Introduction}
\label{sec:introduction}
\IEEEPARstart{S}{ilicon} carbide (SiC) detectors offer significant advantages over conventional silicon detectors, including superior radiation hardness, reduced temperature sensitivity, higher breakdown voltage, lower leakage current, and faster time response. These properties make them attractive for particle physics detection, space radiation monitoring, and medical dosimetry\cite{nava2008, denapoli2022}. Traditional semiconductor detectors typically use all-metal electrodes to collect signals generated by ionizing radiation. However, the presence of metal electrodes is disadvantageous in certain applications, including low-penetration particle detection (e.g., low‑energy soft X‑rays, low‑energy alpha particles, ultraviolet light), transient current technique (TCT) measurements and medical dosimetry\cite{Jiang2026}.  Graphene electrodes exhibit high transmittance to low-energy particles, X-rays, and ultraviolet light, thereby avoiding the signal attenuation and backscattering commonly associated with metal electrodes\cite{ni2017graphene, jabeen2018graphene}. Their atomic thickness reduces the detector dead layer to below the nanometer scale, and in combination with ultrahigh carrier mobility, significantly lowers leakage current and noise \cite{geim2007}. These properties make graphene an ideal electrode material for the active region of detectors, overcoming the drawbacks of metal electrodes in applications such as low-penetration particle detection, transient current technique (TCT) measurements, and medical dosimetry. Furthermore, graphene interlayers can lower the metal/4H-SiC interface barrier and improve the electrical contact performance between SiC and metals \cite{liu2010, jia2021}. More importantly, owing to its high transmittance to X-rays and ultraviolet light, graphene can serve as an electrode material for the active region of detectors to meet the requirements of low-penetration particle detection applications\cite{ni2017graphene, jabeen2018graphene}. Graphene has been shown to improve the signal response time, charge collection stability, and time resolution of radiation detectors.\cite{lopez2024}. A graphene-optimized silicon carbide PIN detector was fabricated and its radiation tolerance under X-ray irradiation of 160 keV was evaluated.

In high-intensity irradiation scenarios such as medical imaging,
security screening, deep-space exploration and scientific research,
people are increasingly emphasizing the radiation hardness of Xray
detectors to ensure long-term stable operation, reduce the risk
of failure, and meet the demands of lower-dose imaging and more
extreme environmental applications. With the development of large scientific facilities such as the Circular Electron Positron Collider (CEPC) and synchrotron applications toward higher brightness, detectors are required to possess high radiation hardness, high time resolution, and high detection accuracy, while maintaining stable operation in intense X-ray background radiation environments\cite{nida2019}.
The development of advanced radiotherapy modalities, such as FLASH radiotherapy with dose rates exceeding 40 Gy/s, requires detectors to maintain accurate dose detection in intense X-ray irradiation environments\cite{nida2019, lopezpaz2024, fleta2024flash}. In space exploration, satellites and probes are continuously bombarded by cosmic rays and solar X-rays, necessitating detectors for navigation, scientific instrumentation, and monitoring electronic systems against single-event effects (SEE) and total ionizing dose (TID) transients. All these requirements demand that detectors possess radiation hardness, high signal-to-noise
ratio, high sensitivity, and fast time response, without degradation due to cumulative effects, while operating stably and providing accurate measurement results. Conventional silicon detectors exposed to high energy X-ray irradiation exhibit pronounced susceptibility to TID effects and displacement damage (DD), which induce a marked rise in leakage current, deterioration of the CCE, and ultimately device failure. Consequently, this severely constrains their deployment in
extreme radiation environments. The significance of studying the
charge collection performance graphene-optimized silicon
carbide PIN detector under high-intensity X-ray irradiation lies in
validating its ability to maintain core functionalities under extreme
radiation environments.

In this work, graphene-optimized ring-electrode (G/RE) 4H-SiC PIN detectors have been fabricated. The current-voltage (I-V) and capacitance-voltage (C-V) electrical characteristics of the detectors were systematically evaluated before and after X-ray irradiation at an energy of 160 keV and doses of 0.1 MGy and 1 MGy. Using TCAD simulations, consistency between simulated and experimental C-V curve is achieved by introducing series resistance and modified
interface parameters in the graphene contact. The CCE and signal rise time of the detector for $\alpha$ and $\beta$ particles before and after irradiation were analyzed, demonstrating the stability of the detector performance. Studying the signal rise time and CCE of $\alpha$ and $\beta$ particles under X-ray irradiation reveals the differential effects of ionization damage on particles with different ionization densities (high-LET  $\alpha$ particles vs. weak-signal $\beta$ particles/MIPs), providing a unified basis for energy/time resolution, contact optimization, and radiation-hardening design of mixed-field detectors.

\section{Detector Fabrication and Irradiation Conditions}
\label{sec:fabrication}
\subsection{Epitaxial Structure and Device Fabrication}
\label{subsec:epitaxial}
Our teem fabricated the G/RE 4H-SiC PIN detectors with a size of 2 mm$\times$2 mm for the first time\cite{Jiang2026}. The G/RE 4H-SiC PIN detector structure includes single-layer graphene, P-type electrode, SiO$_2$ passivation layer, P++ layer, N- epitaxial layer, N-type buffer layer, conductive N-type 4H-SiC substrate, and N-type electrode. The P++ layer has an aluminum ion doping concentration of $2 \times 10^{19}$ cm$^{-3}$ and a thickness of 0.5 $\mu$m. The lightly doped N- epitaxial layer has a nitrogen doping concentration of $5 \times 10^{13}$ cm$^{-3}$ and a thickness of 50 $\mu$m. The N-type buffer layer has a nitrogen doping concentration of $1.2 \times 10^{18}$ cm$^{-3}$ and a thickness of 1 $\mu$m. The device fabrication process mainly involves photolithography, etching, electron beam evaporation, magnetron sputtering, rapid thermal annealing, graphene transfer, and patterning. The epitaxial layer etching depth exceeds 0.5 $\mu$m to ensure complete etching of the P++ layer. The Ni/Ti/Al (50/15/60 nm) electrodes were deposited on the P++ layer and N-type substrate by electron beam evaporation, followed by rapid thermal annealing at 950$^{\circ}$C for 2 minutes to form P-type ohmic contacts. A 500 nm thick SiO$_2$ layer was deposited by plasma-enhanced chemical vapor deposition (PECVD) at 350$^{\circ}$C. Etching of the silicon dioxide layer to open electrode contact windows. The graphene electrode is formed by transferring and reactive ion etching (RIE) of commercial single-layer graphene (SixCarbon Technology).
\begin{figure}[t]
\centering
\includegraphics[width=3.5in]{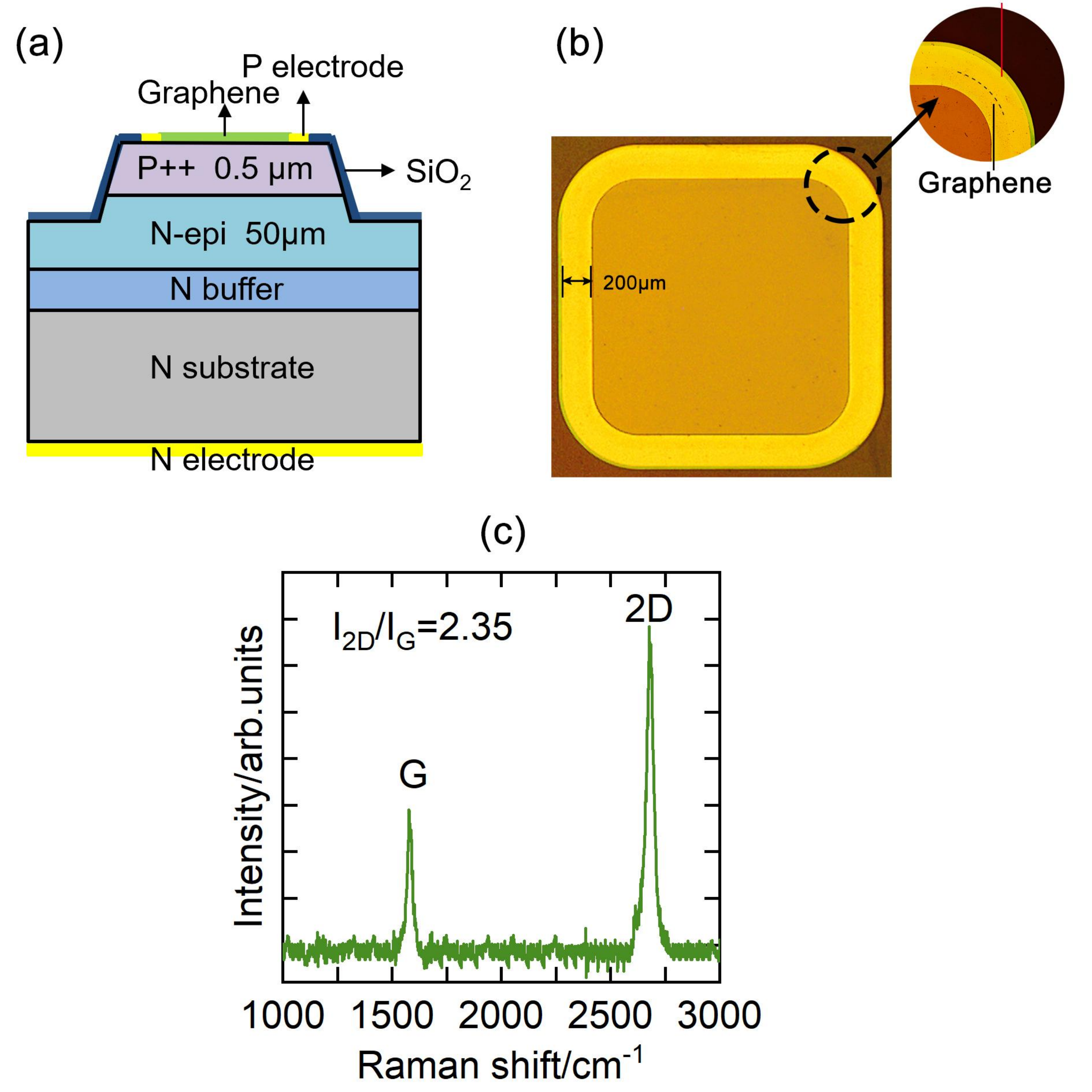}
\caption{(a) Cross-sectional schematic of the G/RE 4H-SiC PIN
radiation detector. (b) Real G/RE 4H-SiC PIN radiation detector
image. (c) Raman spectrum of the G/RE 4H-SiC PIN detector.}
\label{fig:structure}
\end{figure}

The Raman spectrum of graphene was measured at room temperature using a LabRam HR80 laser confocal spectrometer with a laser wavelength of 532 nm, spot radius of 1~$\mu$m, and power of 5 mW. The transferred graphene exhibits distinct G peak (1583 cm$^{-1}$) and 2D peak (2671 cm$^{-1}$). The 2D peak has a perfect single Lorentzian profile, and the I$_{2D}$/I$_G$ ratio is greater than 1.5, confirming the single-layer structure of graphene. The D peak (1340 cm$^{-1}$) represents defect-induced breathing modes, and its low intensity indicates minimal defects introduced during the transfer process.
\subsection{Irradiation Conditions}
\label{subsec:irradiation}
X-ray irradiation was conducted using an X-ray irradiation facility at the Institute of High Energy Physics, Chinese Academy of Sciences. The irradiation energy was 160 keV, and the irradiation doses were 0.1 MGy and 1 MGy, respectively. The irradiation temperature was room temperature, and the devices were unbiased during irradiation.
\section{Electrical Performance Analysis}
\label{sec:electrical}
\subsection{Current-Voltage (I-V) Characteristics}

\begin{figure}[t]
\centering
\includegraphics[width=3.5in]{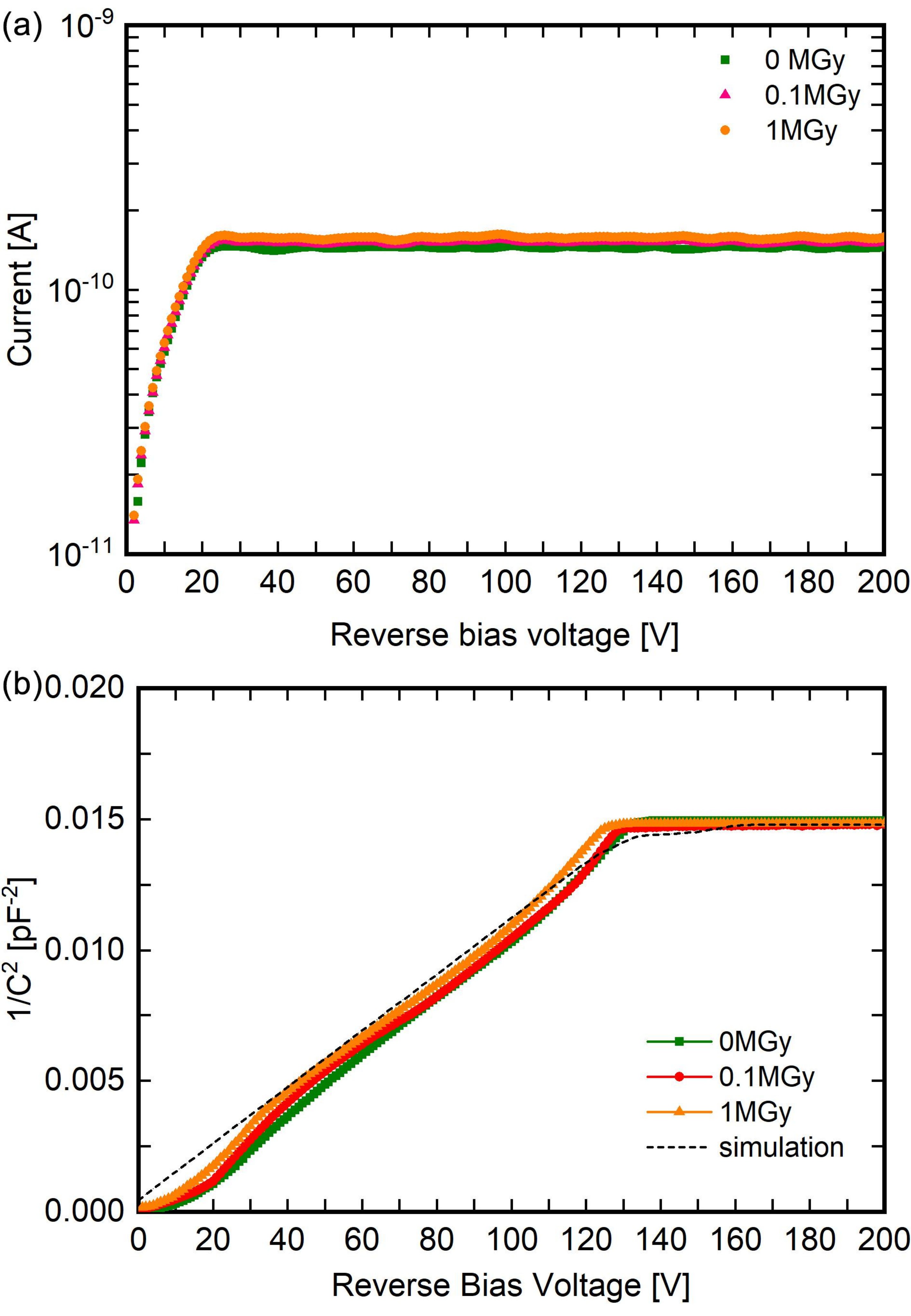}
\caption{(a) I-V characteristics and (b) $1/C^2$-$V$ characteristics of the unirradiated and irradiated G/RE 4H-SiC PIN detectors. In (b), the dashed line represents the TCAD simulation results.}
\label{fig:iv}
\end{figure}

The current-voltage (I-V) characteristics of the devices were measured using a Keithley 2470 source meter on a probe station at room temperature. Fig.~\ref{fig:iv}~(a) shows the I-V characteristics of the G/RE 4H-SiC PIN detectors before and after 160 keV X-ray irradiation. The leakage currents of the unirradiated, 0.1~MGy, and 1~MGy G/RE 4H-SiC PIN detectors at 200~V reverse bias are approximately $1.45 \times 10^{-10}$~A, $1.51 \times 10^{-10}$~A, and $1.57 \times 10^{-10}$~A, respectively. The leakage current shows a slight upward trend with increasing irradiation dose, indicating that the X-ray irradiation caused very little damage to the leakage current of the device. This slight increase in leakage current is consistent with ionization-dominated effects rather than bulk displacement damage. The 160 keV X-ray photons primarily interact via Compton scattering and photoelectric absorption, generating secondary electrons with energies up to 120~keV. However, owing to the large mass difference between electrons and lattice atoms, the maximum kinetic energy transferable to Si or C atoms is only 9 eV and 21 eV, respectively—values well below the displacement threshold energies of 25–35 eV for Si and 20–25 eV for C in SiC\cite{lebedev2004}. Consequently, no significant atomic displacements or bulk trap formation is expected. The observed electrical changes are therefore attributed to ionization‑induced modifications at the graphene/SiC interface and within the graphene layer itself. These modifications manifest as an increase in contact resistance—a phenomenon commonly reported for carbon‑based electrodes exposed to X‑rays\cite{ni2017graphene, jabeen2018graphene}. This behavior is in sharp contrast to that observed under high‑energy proton irradiation, where radiation‑induced deep traps cause substantial carrier removal, leading to a strong suppression of generation current and an orders‑of‑magnitude reduction in leakage current\cite{rafi2020, zhao2015}. Under 1 MGy irradiation, the leakage current of the G/RE 4H‑SiC PIN detector remains below 0.2 nA, demonstrating excellent electrical stability for dosimetry applications.

\subsection{Mott-Schottky (1/C²-V) Characteristics}

Capacitance-voltage (C-V) measurements were performed using a Keysight E4980A precision LCR meter with a high-voltage bias adapter, with a test frequency of 10 kHz and AC signal amplitude of 50 mV. Fig.~\ref{fig:iv}~(b) shows the $1/C^2$-$V$ characteristics of the unirradiated and irradiated G/RE 4H-SiC PIN detectors. The $1/C^2$-$V$ curves clearly remain stable above 120 V, indicating that the detector is fully depleted. The effective doping concentration and depletion
depth of N-epi layer are about $8.08 \times 10^{13}$ cm$^{-3}$ and 42 $\mu$m, respectively. This negligible variation (within experimental error) indicates that 160 keV X-rays produce secondary electrons with insufficient energy to create bulk displacement damage or significant carrier removal in the SiC epitaxial layer, confirming the exceptional radiation hardness of the 4H-SiC bulk material. Fig.~\ref{fig:iv}(b) demonstrate that the effective doping concentration remains stable after irradiation, as confirmed by the consistent slopes in the high-voltage region (100--200 V). The minimal variation demonstrates excellent radiation hardness of the bulk material. The full depletion voltages of the detectors vary slightly, which are due to differences in manufacturing processes rather than non-radiation-induced damage. For instance, even a slight deviation in the thickness of the active layer or in the doping concentration during epitaxial growth can lead to minor differences in the C-V curve. These differences are all within the acceptable manufacturing tolerance range. 

\subsection{TCAD Simulation of Capacitance-Voltage Response}
To elucidate the physical origin of the observed electrical stability and to verify the integrity of the 4H-SiC bulk material under X-ray irradiation, a calibrated TCAD simulation was performed using Synopsys Sentaurus. A one-dimensional model of the G/RE 4H-SiC PIN structure was constructed, incorporating the actual layer stack: P$^{++}$ ($2 \times 10^{19}$~cm$^{-3}$, 0.5~$\mu$m), N$^{-}$ epi-layer (50~$\mu$m), N-buffer ($1.2 \times 10^{18}$~cm$^{-3}$, 1~$\mu$m), and N$^{+}$ substrate ($2 \times 10^{18}$~cm$^{-3}$, 1~$\mu$m). The key challenge lies in modeling the graphene top contact, which exhibits non-ideal behavior due to its van der Waals interface with SiC.

The simulation accounts for this by introducing a voltage-dependent contact resistance and interfacial trap states at the graphene/4H-SiC junction. The doping concentration of the N$^{-}$ epi-layer was initially set to $8.08 \times 10^{13}$~cm$^{-3}$—the value extracted from experimental C-V data—and held constant for all irradiation conditions. Crucially, no radiation-induced bulk traps or changes in doping concentration were included in the model. Despite this, by only increasing the graphene contact resistance from 1200~$\Omega$ (unirradiated) to 1350~$\Omega$ (0.1~MGy) and 2400~$\Omega$ (1~MGy)—reflecting ionization-induced degradation in the graphene layer—the simulated $1/C^2$-$V$ characteristics (dashed curves in Fig.~\ref{fig:iv}(b)) reproduce the experimental measurements with high fidelity across the entire bias range.

The excellent agreement between simulation and experiment, achieved without modifying the bulk properties of 4H-SiC, provides strong evidence that the slight increase in series resistance and minor CCE degradation originate solely from the graphene electrode and its interface, not from displacement damage or carrier removal in the semiconductor bulk. This further confirms that 160~keV X-ray irradiation induces predominantly ionization effects, leaving the exceptional radiation hardness of 4H-SiC fully intact.

\subsection{Effective Doping Concentration and Depletion Depth Characteristics}
The effective doping concentration and depletion depth can be calculated using the $1/C^2$-$V$ curve. Fig. 3 presents the effective doping concentration profiles
extracted from the C-V data using the differential method. A one-dimensional model of a 4H-SiC PIN diode was established using Sentaurus TCAD software to aid in understanding the internal structure of the detector. The thickness of the P++ layer is set to 0.5 $\mu$m with a doping concentration of $2 \times 10^{19}$ cm$^{-3}$. The lightly doped N- epitaxial layer has a thickness of 50 $\mu$m  and a doping concentration of $8.08 \times 10^{13}$ cm$^{-3}$, which was obtained from CV curve calculation. The thickness of an N-type buffer layer is set to 1$\mu$m with a doping concentration of $1.2 \times 10^{18}$ cm$^{-3}$. The thickness of the conductive N-type 4H-SiC substrate is set to 1$\mu$m with a doping concentration of $2 \times 10^{18}$ cm$^{-3}$. The C-V characteristics were simulated using the SDevice module, with trap models including TrapRecombination, TrapOccupation, and TrapDOS employed to account for contact effects at the SiC/graphene boundary. Adjusting the contact resistance between graphene and silicon carbide, the simulation results agree with the experimental results. This confirms that the observed electrical characteristics originate from contact effects rather than bulk material degradation.

\begin{figure}[t]
\centering
\includegraphics[width=3.5in]{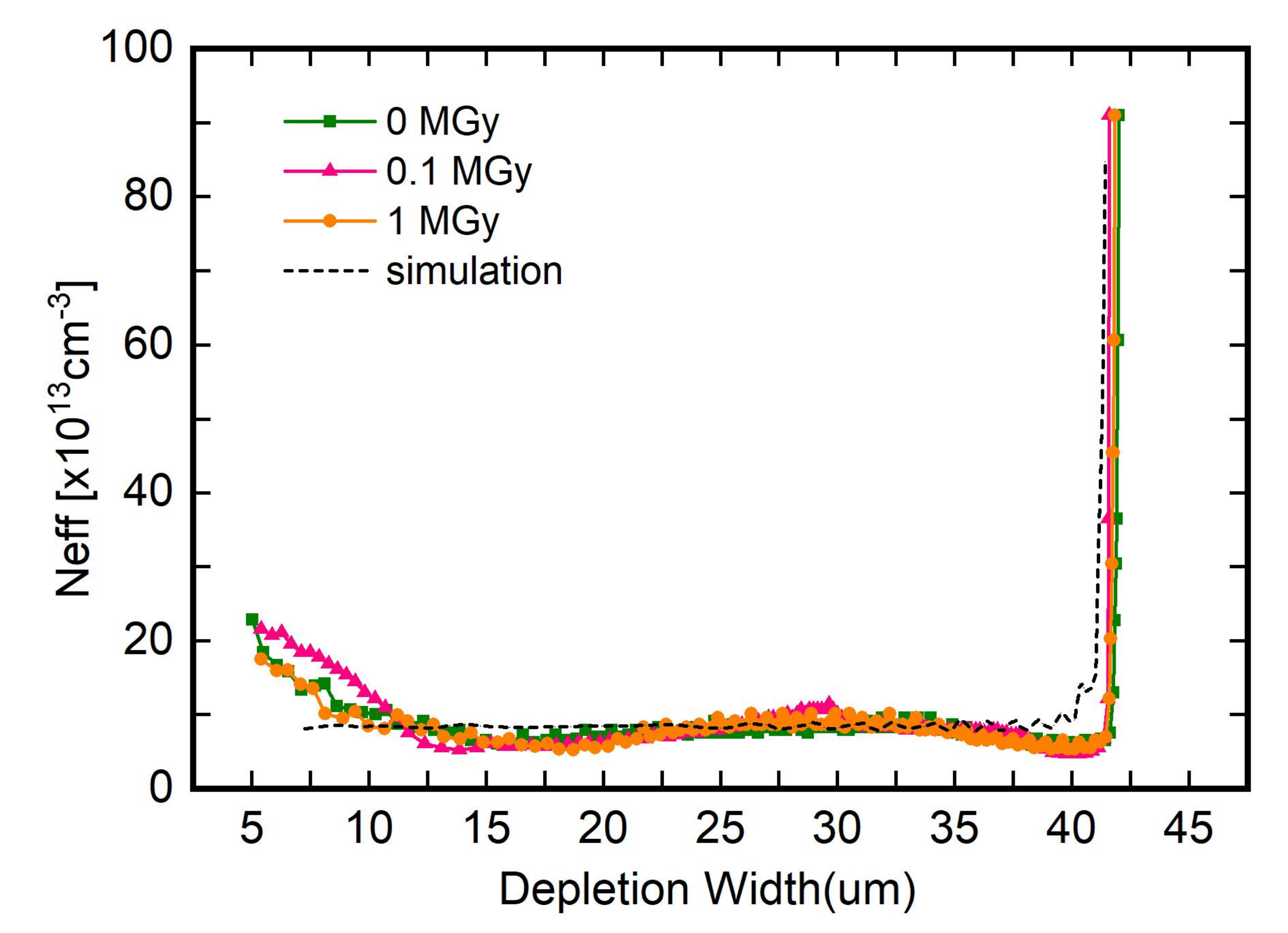}
\caption{Effective doping concentration profiles as a function of depletion depth for unirradiated and irradiated detectors. The dashed line indicates the designed doping concentration of the N-epitaxial layer.}
\label{fig:doping_profile}
\end{figure}

Fig.~\ref{fig:doping_profile} presents the effective doping concentration profiles extracted from the C-V data using the differential method.
 All samples exhibit a distinct plateau region between approximately 10~$\mu$m and 40~$\mu$m, with a nearly constant doping concentration of approximately $8.08 \times 10^{13}$ cm$^{-3}$, which matches well with the designed doping level of the N-epitaxial layer. The sharp rise beyond 4~$\mu$m indicates the transition to the highly doped N-type buffer/substrate interface. Notably, the profiles remain essentially unchanged after irradiation and show excellent agreement with the designed doping level (dashed line), further confirming the stability of the bulk material.
 
\section{Rise Time and Charge Collection Performance Analysis}

\subsection{Experimental setup}

\begin{figure}[htbp]
\centering
\includegraphics[width=3.5in]{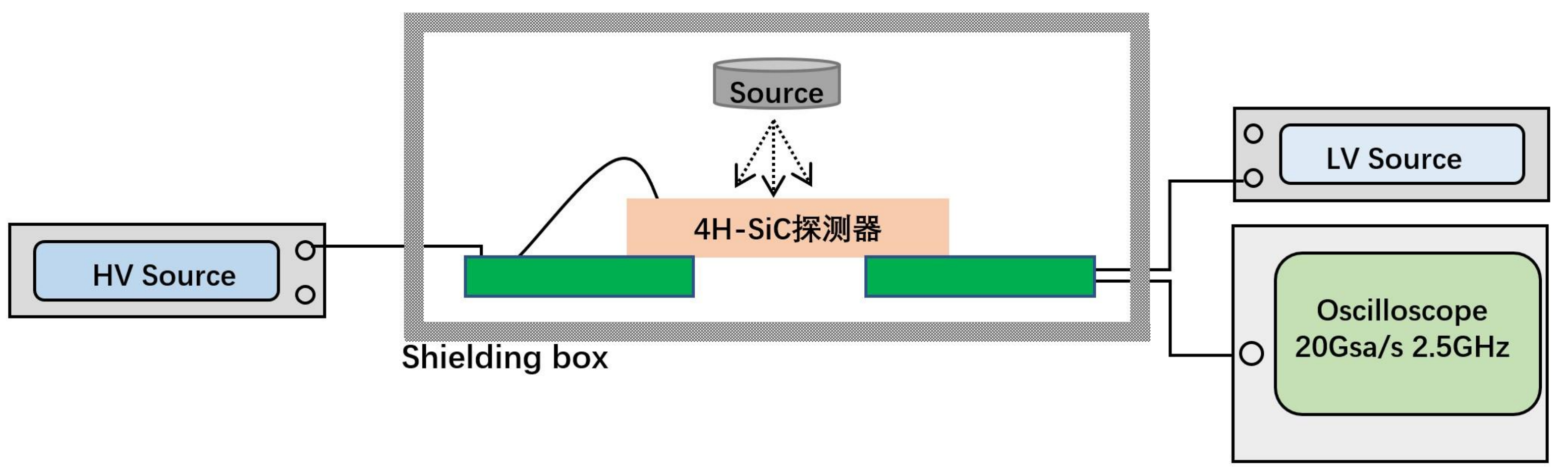}
\caption{Experimental setup for charge collection efficiency of a $^{241}$Am or $^{90}$Sr source.}
\label{fig:setup}
\end{figure}

The charge collection performance test system includes an $^{241}$Am or $^{90}$Sr radioactive source, G/RE 4H-SiC detector, single-channel electronic readout board, low-voltage power supply (GPD-3303), high-voltage power supply (Keithley 2470), and oscilloscope (MSO64, Tektronix 2.5 GHz), as shown in Fig.~\ref{fig:setup}. The 4H-SiC detector is mounted on the electronic readout board using conductive adhesive, with the detector pad electrodes connected to the readout board. The high-voltage power supply provides reverse bias to the detector, and the low-voltage power supply powers the single-channel electronic readout board. When alpha particles emitted by the radioactive source pass through the device, they generate current signals, which are amplified by the electronic amplifier and converted to voltage signals. The oscilloscope acquires the pulse waveforms, and charge collection information is obtained by integrating the pulse waveforms.
\subsection{Rise Time and Charge Collection Performance for $\alpha$ particles}

The core importance of studying the rise time and CCE of a G/RE 4H-SiC PIN detector for alpha particles under X-ray irradiation lies in revealing how ionization damage affects subsequent or simultaneous alpha particle detection performance. This provides key experimental evidence for mixed-field detector design, radiation hardening assessment, and high-reliability applications (e.g., nuclear waste characterization, fusion device diagnostics).

\subsubsection{Rise Time for $\alpha$ particles}

\begin{figure}[htbp]
\centering
\includegraphics[width=3.5in]{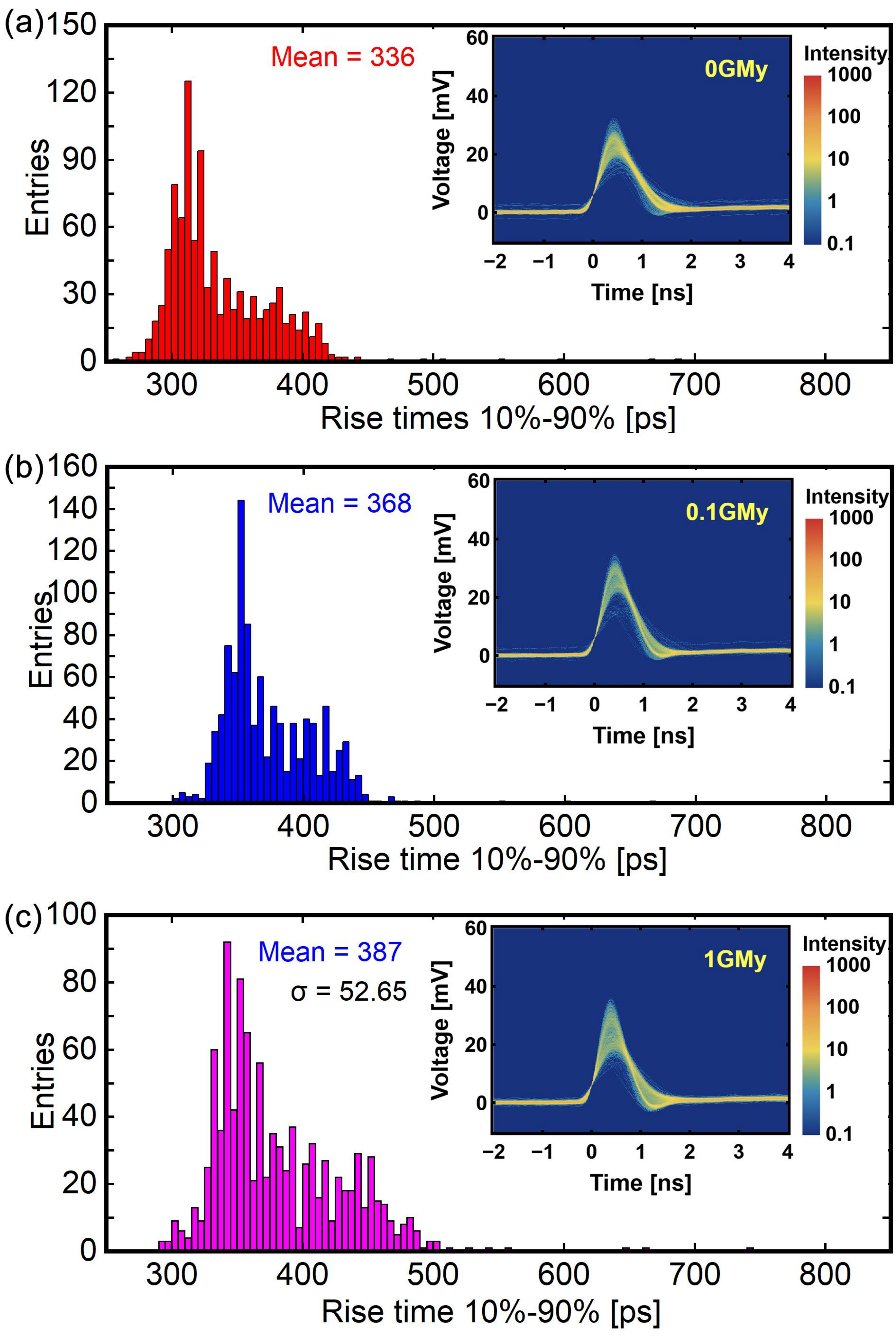}
\caption{Signal waveforms and rise time distributions at 200~V for $\alpha$ particles: (a) 0 MGy X-ray irradiated G/RE 4H-SiC PIN. (b) 0.1 MGy X-ray irradiated G/RE 4H-SiC PIN. (c) 1 MGy X-ray irradiated G/RE 4H-SiC PIN.}
\label{fig:risetime}
\end{figure}

Fig.~\ref{fig:risetime} shows the rise time distribution of the G/RE 4H-SiC PIN detector before and after 160-keV X-ray irradiation. The rise times of the unirradiated, 0.1 MGy irradiated, and 1 MGy irradiated detectors are approximately 336~ps, 368~ps, and 387~ps, respectively. The G/RE 4H-SiC PIN detector signal at an irradiation dose of 0.1 MGy maintains a sharp pulse profile with minimal tailing, while the detector signal at an irradiation dose of 1 MGy exhibits obvious pulse broadening and significant long tails after 2 ns. After 0.1 MGy and 1 MGy X-ray irradiation, the signal rise time for $\alpha$ particles of the detector increased slightly by 9.5\% and 15.1\%, respectively. The reduced peak amplitude and increased waveform dispersion of the 1 MGy device not only indicate quantitative charge loss but also reflect increased statistical fluctuations during the collection process. 

\label{subsec:cce}

Irradiation-induced increase in the series resistance of the graphene layer and modification of the carrier transport properties at the SiC/graphene contact are the two main causes of the detector’s signal rise time\cite{pellegrini2016, lange2010}. Nevertheless, the detector maintains a sub-nanosecond time response even after 1 MGy high-dose X-ray irradiation, demonstrating the good performance of the graphene-optimized SiC detector in high-radiation environments.

\subsubsection{Charge Collection Performance for $\alpha$ particles}

\begin{figure}[t]
\centering
\includegraphics[width=3.5in]{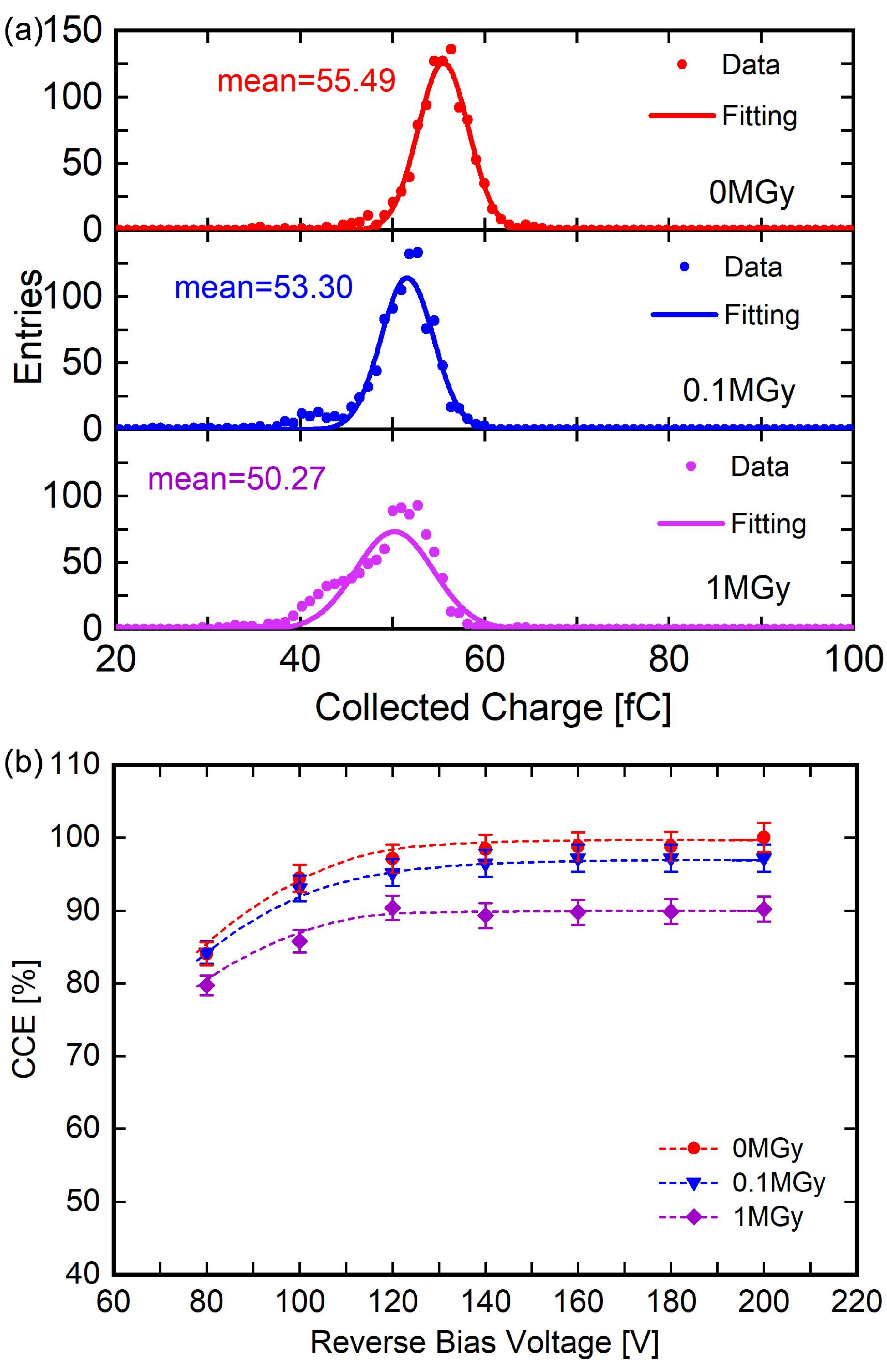}

\caption{Charge collection performance 200~V for $\alpha$ particles: (a) Landau fit of the collected charge spectrum, with MPV indicating characteristic charge of detectors. (b) CCE versus reverse bias for the 0 MGy, 0.1 MGy and 1 MGy X-ray irradiated G/RE 4H-SiC PIN.}
\label{fig:cce}
\end{figure}

The $^{241}$Am source decays to release $\alpha$ particles with an energy of 5.486 MeV. It can be fully absorbed by detectors with 50~$\mu$m epitaxial layer\cite{gorin2007, mozer2018}. The $\alpha$ particles can produce electron-hole pairs after it passes through the depleted layers. Fig.~\ref{fig:cce}~(a) shows the collected charge at 200 V bias of the G/RE 4H-SiC PIN detector before and after X-ray irradiation. The charge collection spectra for alpha particles from the detector before and after irradiation exhibit good Gaussian distributions. The collected charges for the unirradiated, 0.1 MGy, and 1 MGy X‑ray irradiated samples are 55.49~fC, 53.3~fC, and 50.27~fC, respectively. The CCE of the unirradiated G/RE 4H-SiC PIN detector was defined as 100\% @ 200 V. The CCEs of the unirradiated, 0.1 MGy and 1MGy G/RE 4H-SiC PIN are 97.2\% and 90.0\% at 200 V, shown in Fig.~5~(b). After X-ray irradiation with doses of 0.1 MGy and 1 MGy, the charge collection efficiency decreased by only 3\% and 10\%, respectively. The charge collection efficiency after irradiation exhibits good stability, indicating that the device has resistance to X-ray irradiation.

Irradiation leads to an increase in the series resistance of the graphene contact layer and modifications of the electrical properties at the SiC/graphene interface. These effects reduce the effective electric field in the active region and hinder carrier extraction, thereby degrading the charge collection efficiency, prolonging the signal rise time, and worsening the time resolution. Meanwhile, the SiC bulk material itself possesses high radiation hardness, and its carrier transport properties remain largely unaffected after irradiation\cite{spieler2005}. Consequently, the degradation of the rise time is primarily governed by the graphene contact layer and its interface with SiC\cite{li2018, zhao2015}.

\subsection{Rise Time and Charge Collection Performance for $\beta$ particles}

The importance of studying signal rise time and CCE of a G/RE 4H-SiC PIN detector for $\beta$ particles under X-ray irradiation lies in quantitatively revealing, from the perspective of weak ionization signal extraction, the degradation behavior of cumulative ionization damage on continuous-energy-spectrum beta particles—ranging from the low-energy, high energy-loss region to the high-energy MIPs region. This provides indispensable experimental data and design guidelines for the accuracy of beta radiation monitoring in mixed fields, maximum count rate, energy spectrum fidelity, and, furthermore, for the radiation hardness reliability of timing and position detection using high-energy beta particles as a representative of MIPs.

\subsubsection{Rise Time for  $\beta$ particles}

\begin{figure}[htbp]
\centering
\includegraphics[width=3.5in]{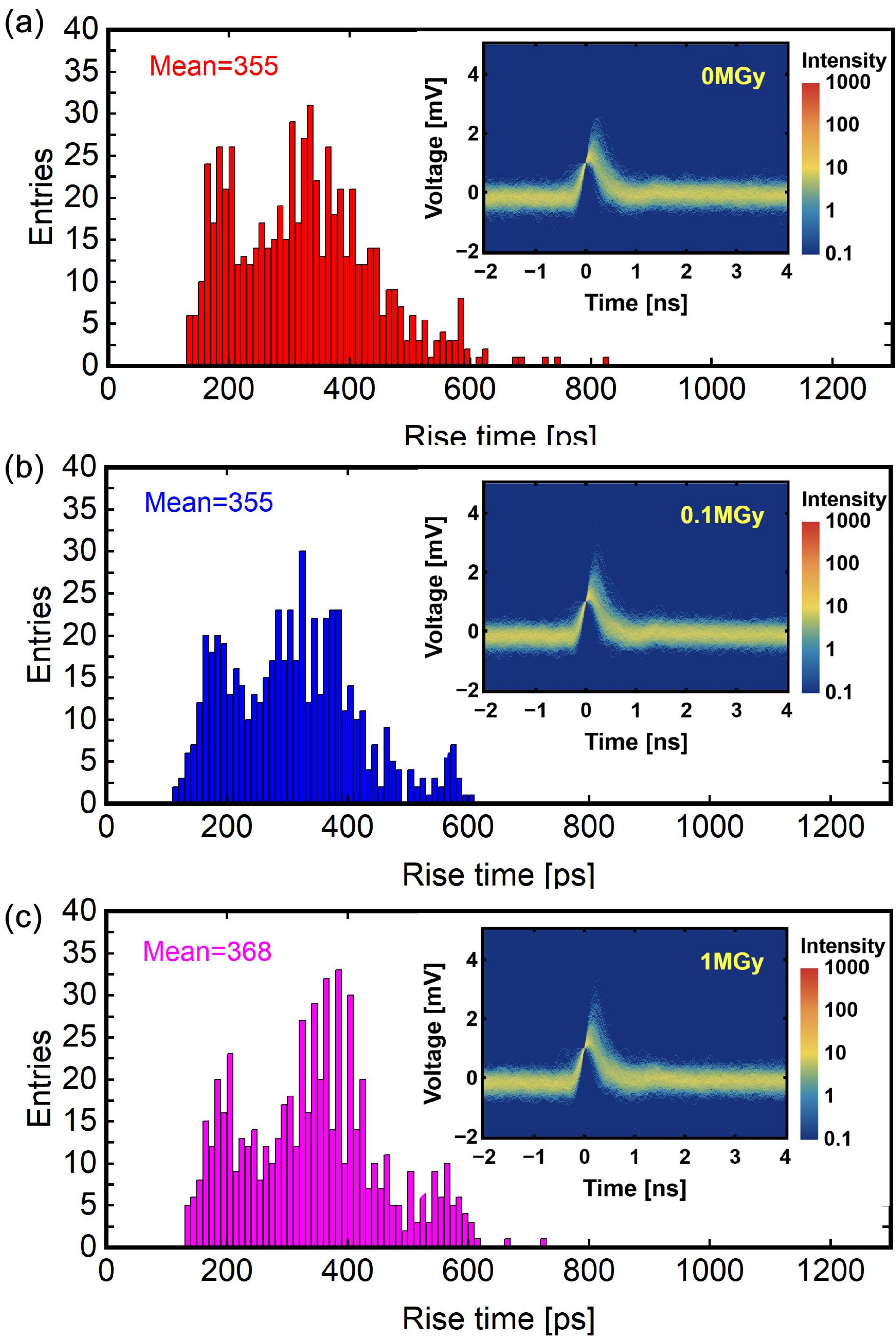}
\caption{Signal waveforms and rise time distributions at 200~V for $\beta$ particles: (a) 0 MGy X-ray irradiated G/RE 4H-SiC PIN. (b) 0.1 MGy X-ray irradiated G/RE 4H-SiC PIN. (c) 1 MGy X-ray irradiated G/RE 4H-SiC PIN.}
\label{fig:risetime_beta}
\end{figure}

Fig.~\ref{fig:risetime_beta} shows the rise time distribution of the detector under $\beta$-irradiation. The signal rise times of the unirradiated, 0.1 MGy X-ray irradiated, and 1 MGy X-ray irradiated G/RE 4H-SiC PIN detectors are approximately 355~ps, 355~ps, and 368~ps, respectively. The G/RE 4H-SiC PIN detector signal at an irradiation dose of 0.1 MGy maintains a sharp pulse profile with minimal tailing, while the detector signal at an irradiation dose of 1 MGy exhibits obvious pulse broadening and significant long tails after 2 ns. After 0.1 MGy X-ray irradiation, the signal rise time for $\beta$ particles showed almost no change. However, after 1 MGy X-ray irradiation, the signal rise time for $\beta$ particles increased slightly by 16.4\%. The reduction in peak amplitude and the increase in waveform dispersion confirm that the performance degradation is not particle-specific. It originates from intrinsic changes in the detector's electrical properties, primarily the increase in series resistance of the graphene contact layer. TCAD simulations validate this conclusion.

\subsubsection{Charge Collection Performance for $\beta$ particles}

\begin{figure}[htbp]
\centering
\includegraphics[width=3.5in]{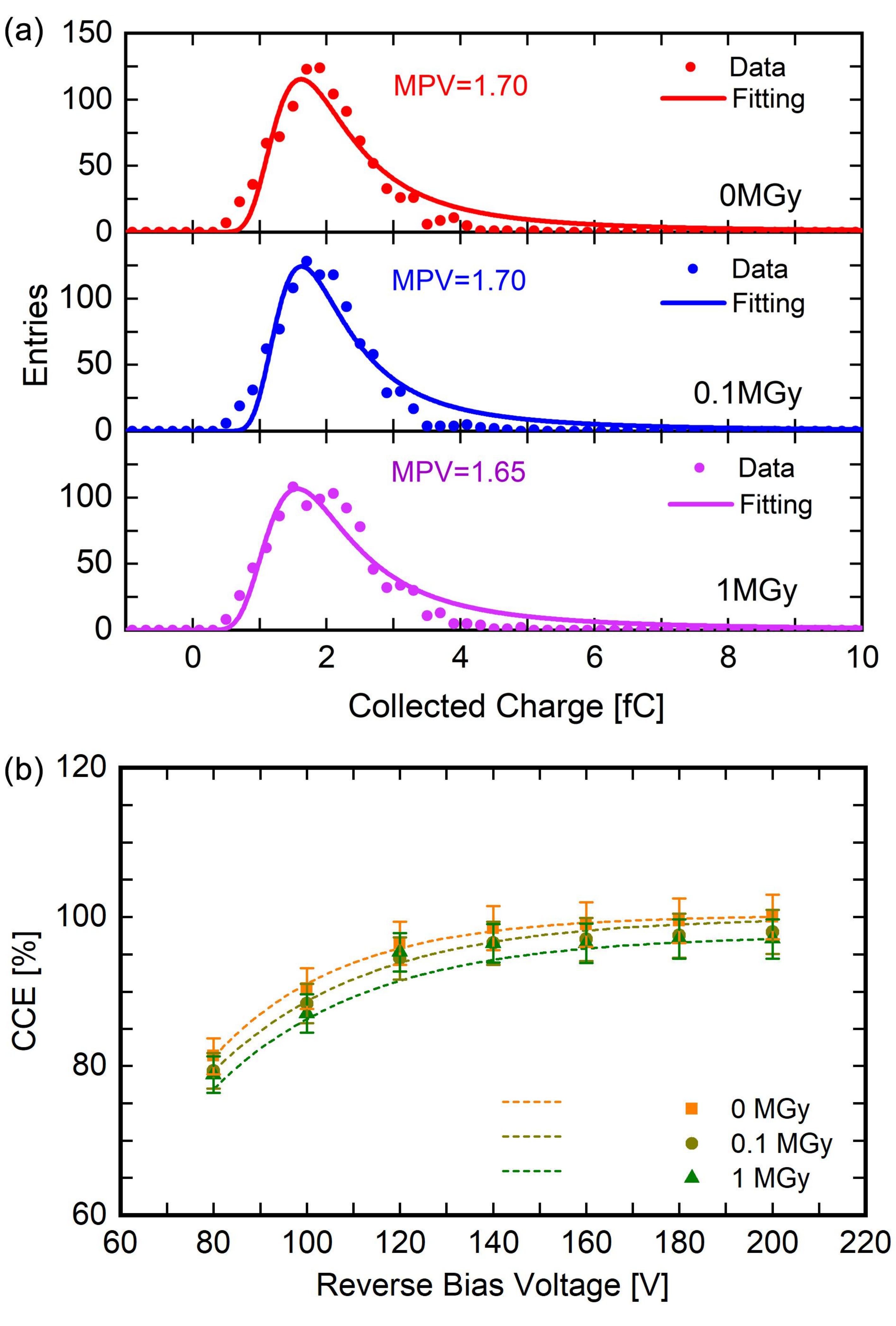}
\caption{Charge collection performance 200~V for $\beta$ particles: (a) Landau fit of the collected charge spectrum, with MPV indicating characteristic charge of detectors. (b) CCE versus reverse bias for the 0 MGy, 0.1 MGy and 1 MGy X-ray irradiated G/RE 4H-SiC PIN.}
\label{fig:cce_beta}
\end{figure}

The$^{90}$Sr source decays to release $\beta$ particles with a maximum energy of 0.546 MeV. Fig.~\ref{fig:cce_beta} shows the collected charge at 200 V bias of the G/RE 4H-SiC PIN detector before and after X-ray irradiation. The charge collection spectra for $\beta$ particles from the G/RE 4H-SiC PIN detector before and after irradiation exhibit good  Landau distributions. The collected charges for the unirradiated, 0.1 MGy, and 1 MGy X‑ray irradiated samples are 1.70~fC, 1.70~fC, and 1.65~fC, respectively. The CCE of the unirradiated G/RE 4H-SiC PIN detector was defined as 100\% @ 200 V. The CCEs of the unirradiated, 0.1 MGy and 1MGy G/RE 4H-SiC PIN are 97\% and 90\% at 200 V, shown in Fig.~\ref{fig:cce_beta}(b). 

The unirradiated device exhibits a well-defined spectral shape with a clear Compton edge, and achieves near full charge collection at 120~V bias. Using the average collected charge of the unirradiated detector at 200~V as the 100\% baseline, the charge collection efficiency shows systematic degradation with increasing irradiation dose: the 0.1~MGy irradiated sample achieves approximately 100.0\% CCE at 200~V, while the 1~MGy sample decreases to 97.0\%. Similar to the $\alpha$-particle results, the CCE curve of the 1~MGy sample exhibits premature saturation above 120~V, indicating a fundamental limitation in charge collection that cannot be overcome by increasing bias. The Fig.~\ref{fig:cce_beta}(a) shows the unirradiated sample presents a smooth distribution peaking at 1.70~fC, while the 1~MGy sample's peak shifts left to 1.65~fC, corresponding to an $\sim$12\% reduction in collected charge. The CCE is calculated based on the mean energy deposition of $\beta$-particles in 4H-SiC, with an average generation energy of $\sim$13.9~eV per electron-hole pair \cite{gorin2007}. Given the continuous energy spectrum and longer range of $\beta$-particles (up to $\sim$1~mm in SiC), the active region is fully sensitive, and the observed CCE reflects the integrated collection efficiency across the entire epitaxial layer.

\section{Conclusion}
\label{sec:conclusion}

A novel graphene-optimized silicon carbide PIN detector was fabricated. Its electrical properties, charge collection performance and signal rise time were evaluated under non‑irradiated conditions and under X‑ray irradiation with an energy of 160 keV at doses of 0.1 MGy and 1 MGy. The leakage currents of the detectors under non‑irradiated, 0.1 MGy, and 1 MGy irradiation conditions are approximately $1.45 \times 10^{-10}$~A, $1.51 \times 10^{-10}$~A, and $1.57 \times 10^{-10}$~A, respectively. The effective doping concentration of the detector is approximately $8.08 \times 10^{13}$ cm$^{-3}$ before and after irradiation, with no significant change. Using TCAD simulations, consistency between simulated and experimental C-V curve is achieved by introducing series resistance and modified interface parameters in the graphene contact. 
The rise times of the $\alpha$ particles signal detected by the detector under unirradiated, 0.1 MGy, and 1 MGy X-ray irradiation conditions are 336 ps, 368 ps, and 387 ps, respectively. The rise times of the $\beta$ particles signal detected by the detector under unirradiated, 0.1 MGy, and 1 MGy X-ray irradiation conditions are 342~ps, 375~ps, and 398~ps, respectively. After 0.1 MGy and 1 MGy X‑ray irradiation, the charge collection efficiencies of the detector for $\alpha$ particles are 97.2\% and 90.0\%, respectively; for $\beta$ particles, they are 100.0\% and 97.0\%, respectively. A graphene-optimized silicon carbide PIN detector model was established using TCAD to simulate the effective doping concentration. The model confirms that after 1 MGy of 160 keV X‑ray irradiation, the 4H‑SiC material maintains its electrical integrity. The slight degradation in charge collection efficiency and rise time is fully attributed to the increase in graphene contact resistance caused by ionization effects, rather than to bulk material damage. Experiments confirm that 160 keV X-ray irradiation may not cause significant displacement damage in the 4H-SiC, and the minor performance degradation may be attributed to ionization-induced changes in the graphene electrode.

The detector exhibits excellent charge collection performance and fast time response. These results demonstrate stable performance under extreme X-ray exposure, highlighting the detectors potential for radiation-hard applications in high-energy physics, space missions, and nuclear reactor monitoring.

Our team has successfully fabricated graphene-optimized SiC PiN and SiC LGAD detectors. In addition, we will continue to advance the research and development of next-generation devices, including SiC AC-LGAD and SiC BJT. The SiC AC-LGAD enables simultaneous measurement of submillimeter spatial
resolution and picosecond temporal resolution; the SiC BJT explores
potential applications in detector readout electronics and signal amplification. The development of these novel devices will further enrich the SiC detector technology portfolio and promote their engineering applications in four-dimensional tracking, time-resolved imaging, and extreme radiation environments.

\bibliographystyle{IEEEtran}
\bibliography{references}
\end{document}